\documentclass[12pt,preprint]{aastex}
\usepackage{amsmath}
\usepackage{graphicx}

\newcommand{\eqb}{\begin{equation}}
\newcommand{\eqe}{\end{equation}}
\begin{document}

\title{A new mechanism for dissipation of alternating fields in Poynting dominated outflows}

\author{Yuri Lyubarsky}
 \affil{Physics Department, Ben-Gurion University, P.O.B. 653, Beer-Sheva 84105, Israel}
\begin{abstract}
Reconnection of alternating magnetic fields is an important energy transformation mechanism in Poynting dominated outflows. We show that the reconnection is facilitated by the Kruskal-Schwarzschild instability of current sheets separating the oppositely directed fields. This instability, which is a magnetic counterpart of the Rayleigh-Taylor instability, develops if the flow is accelerated. Then the plasma drips out of the current sheet providing conditions for rapid reconnection. Since the magnetic dissipation leads to the flow acceleration, the process is self-sustaining. In pulsar winds, this process could barely compete with the earlier proposed dissipation mechanisms. However, the novel mechanism turns out to be very efficient at AGN and GRB conditions.
\end{abstract}


\maketitle

\section{Introduction}

It is now widely agreed that relativistic outflows are  launched hydromagnetically. The commonly invoked scenario is that a strong magnetic field in a rapidly rotating compact object serves to convert the rotational energy into the energy of the outflow (see, e.g., the review by Spruit (2010)). The main  advantage of the magnetic launch mechanism is that the magnetic field lines, like driving belts, could transfer the rotational energy to a low density periphery of the central engine thus forming a baryon pure outflow with the energy per particle significantly exceeding the rest mass energy. Such outflows potentially capable of reaching relativistic velocities. Since the energy is transported, at least initially, in the form of the Poynting flux, the question arises of how and where the electromagnetic energy is eventually transformed to the plasma energy. An important point is that the flow could be considered as true matter dominated only when the electro-magnetic energy falls well below the equipartition level because only then the efficient shock dissipation becomes possible (Kennel \& Coroniti 1984).

In the scope of ideal magnetohydrodynamics (MHD), the energy could in principle be transferred to the plasma via gradual acceleration by electromagnetic stresses. The problem is that in relativistic flows, the magnetic and electric forces nearly compensate each other so that the flow is nearly ballistic. It turns out that
unconfined flows stops accelerating when still being highly Poynting dominated.
A more efficient acceleration regime is achieved if the flow is collimated by an external medium. Such a configuration arises naturally in gamma-ray bursts (GRBs), where the relativistic jet from the collapsing stellar core pushes its way through the stellar envelope. In accreting systems, the
outflow from the rotating black hole could be collimated
by the pressure of a slow (and generally magnetized) wind from the outer parts of the accretion disc. However, a systematic study  shows (Komissarov et al. 2007, 2009; Tchekhovskoy et al. 2008, 2010; Lyubarsky 2009) that even though externally confined flows could in principle be accelerated till the equipartition between the kinetic and magnetic energy, the conditions of that are rather restrictive. Moreover, the true matter dominated stage
could in any case be achieved only at exponentially large distances (Lyubarsky 2010).

An efficient conversion of the electro-magnetic into the plasma energy was claimed for impulsive flows (Granot et al. 2010,; Lyutikov 2010a,b). In a  freely expanding shell, the magnetic energy is indeed eventually transformed into the kinetic energy. However, Levinson (2010) demonstrated that in the real world, it is difficult to provide conditions for the free expansion because even a tenuous ambient medium hampers the expansion.
Intermittent ejection of small sub-shells could hardly help because the shells merge while still highly magnetized unless the distance between them is much larger than their width.

The problem of the energy conversion could be resolved if one abandons the assumption of no dissipation. However, one has to find a reliable dissipation mechanism in highly conductive space plasma.
One of the ideas is that some sort of MHD instability destroys the regular structure of the magnetic field thus triggering the anomalous dissipation. The kink instability is the best candidate (Begelman 1998; Giannios \& Spruit 2006).
However, any global MHD instability could develop only if the proper Alfven crossing time is less than the propagation time, which implies $\theta\gamma<1$, where $\gamma$ is the jet Lorentz factor, $\theta$ the opening angle. This condition is rather restrictive; for example it could hardly be fulfilled in GRBs (e.g. Tchekhovskoy et al. 2010). Even if the kink instability does develop, it is still not clear whether the flow is disrupted or just helically distorted. The impact of the instability on the jet structure could be studied only with three-dimensional numerical simulations. Till now only a few attempts was made, the results still being controversial (Moll \& Spruit 2008; (McKinney \& Blandford 2008).

Magnetic dissipation could occur if a small-scale structure with oppositely directed fields preexisted in the flow. In pulsar winds, such a structure arises naturally because the magnetic field in the equatorial belt of the flow changes sign every half of period. It is the annihilation of the oppositely directed fields that provides the main energy conversion mechanisms in pulsar winds (Coroniti 1990; Lyubarsky \& Kirk 2001; Kirk \& Skj{\ae}raasen 2003; Petri \& Lyubarsky 2007; Zenitani \& Hoshino 2007; see also the review by Kirk et al. (2009)). The outflow with alternating fields could  in principle arise also in accreting systems if the magnetic field in the central engine changes sign.
Then independently of the exact field configuration at the launch site, the flow expansion ensures that in the far zone,  the overall magnetic structure is that of the "striped wind" with stripes of oppositely directed azimuthal fields separated by current sheets. The width of the stripes is small as compared with the scale of the flow therefore locally the structure of the flow is very simple: plane current sheets separate domains with oppositely directed magnetic fields. Within the current sheets, the magnetic field changes sign and the pressure balance is maintained by the thermal pressure of the plasma.

Efficient annihilation of oppositely directed magnetic fields across the current sheet implies some sort of strong anomalous resistivity. The necessary condition is that the drift velocity of the current carriers becomes large enough, which is equivalent to the condition that the particle Larmor radius is not small as compared with the thickness of the sheet. Since the current in the sheets decays as $r^{-1}$ whereas the plasma density drops down as $r^{-2}$, the anomalous dissipation due to the charge starvation should inevitably occur at a large enough distance from the source. In pulsar winds, the plasma density is extremely low so that the alternating fields do dissipate in the far zone of the wind or at the wind termination shock (Kirk \& Skj{\ae}raasen 2003; Petri \& Lyubarsky 2007; Zenitani \& Hoshino 2007). However in accreting systems, the outflow is expected to be heavily loaded by the plasma from the accretion disk (and in GRBs, also by the plasma from the progenitor star (Levinson \& Eichler 2003)) therefore the charge starvation conditions are achieved only at unreasonably large distances.

Spruit et al. (2001) postulated that in GRB outflows, the alternating fields annihilate
with a rate $\sim 0.1$ of the speed of light. The model based on this assumption is capable of explaining many essential features of GRBs (Drenkhahn 2002; Drenkhahn \& Spruit 2002; Giannios 2006, 2008; Giannios \& Spruit 2007).
The problem is that the reconnection rate of the order of 10\% from the Alfven velocity is the maximal possible reconnection rate, which is achieved only at special conditions (see review by Yamada et al. 2010). For example, there is practically no reconnection across the heliospheric current sheet, which is a prototype of large-scale current sheets in any astrophysical outflow.

In this Letter, we propose a mechanism for fast reconnection of the alternating fields in Poynting dominated outflows. The basic idea is the following. If the flow is accelerated (or decelerated), an effective gravity force appears in the proper plasma frame so that the hot plasma within the current sheet is supported against the gravity force by the magnetic pressure. Such a configuration is known to be unstable with respect to the Kruskal-Schwarzschild instability, which is a magnetic counterpart of the Rayleigh-Taylor instability. Under the influence of the effective gravity force, the plasma could drip down between the magnetic field lines (e.g. Infeld (1989)) so that bridges between trickles thin out until the current sheet tears (Fig. 1). Thus the instability facilitates the reconnection. An important point is that the magnetic dissipation forces the flow to accelerate (Lyubarsky \& Kirk 2001; Drenkhahn 2002; Drenkhahn \& Spruit 2002; Kirk \& Skj{\ae}raasen 2003) therefore the process is self-sustaining:  the acceleration is an aid to the reconnection whereas the reconnection promotes the acceleration. In this Letter, we show that the alternating fields could be completely dissipated by this mechanism and find the characteristic dissipation scale.

\begin{figure*}
\includegraphics[scale=0.8]{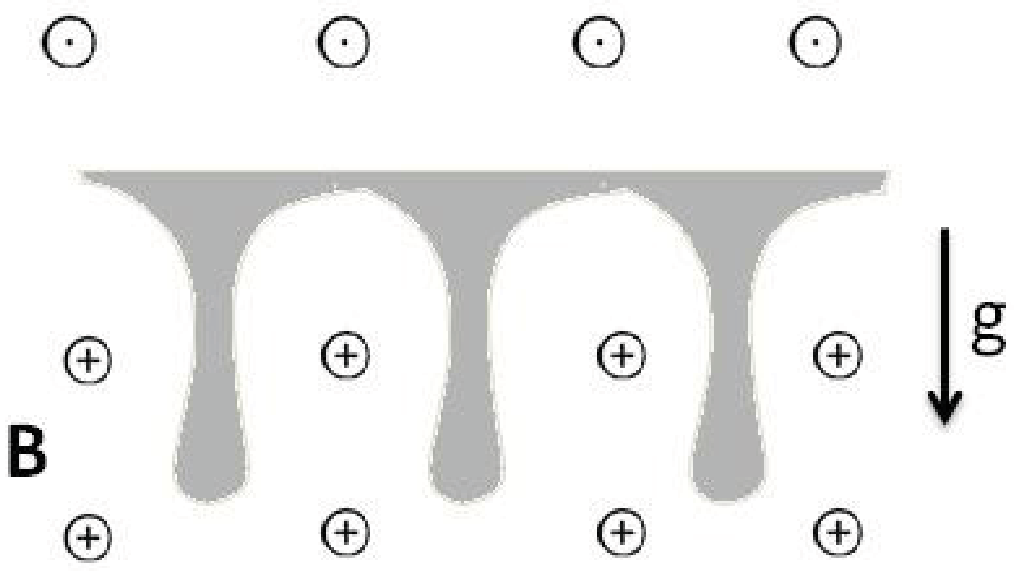}
\caption{The development of the Kruskal-Schwarzschild instability in accelerating flows. In the presence of an effective gravity force, the plasma drips out of the current sheet that separates the oppositely directed magnetic fields. This promotes the magnetic reconnection.}
\end{figure*}

\section{Kruskal-Schwarzschild instability of the current sheet}

Let us consider a hot, nonmagnetized plasma slab, $0<z<\Delta$, separating domains with oppositely directed magnetic fields. Let the field be directed along the $y$ axis. We assume that this structure is accelerated in $z$ direction; then an effective downward gravity force arises in the proper frame of the flow. In the flow frame, the continuity and Euler equations for slow motions are written as
\begin{eqnarray}
\frac{\partial n}{\partial t}+\nabla\cdot n\mathbf{v}=0;\label{cont}\\
h\frac{\partial \mathbf{v}}{\partial t}
=-c^2\nabla p+h\mathbf{g};\label{Euler}
\end{eqnarray}
where $g$ is an effective gravity,
\eqb
h=mnc^2+\frac{\Gamma}{\Gamma-1}p
\label{enthalpy}\eqe
the specific enthalpy, $\Gamma$ the adiabatic index.

In the strongly magnetized domain, the flux freezing condition ties the plasma and the magnetic field together. Therefore when considering flat motions perpendicular to the magnetic field ($v_y=0$), one can take $B/n=\it const$. This implies the effective equation of state $p=B^2/8\pi\propto n^2$; $h=B^2/4\pi=2p$ so that one can describe the magnetized domain by Eqs. (\ref{cont}), (\ref{Euler}) and (\ref{enthalpy}) with $m=0$ and $\Gamma=2$.

We consider evolution of small perturbations, $h=h_0+h_1$, $p=p_0+p_1$, $n=n_0+n_1$, where the unperturbed state satisfies the equilibrium equation
 \eqb
 c^2\frac{dp_0}{dz}=-gh_0.
 \label{equil}\eqe
We assume that the initial state is nearly homogeneous, which implies
 \eqb
 gh_0 \Delta\ll c^2p_0.
 \eqe
Both in the magnetized domain and in the relativistically hot slab, $p\sim h$ therefore
this condition could be written as
 \eqb
 \Delta\ll c^2/g.
 \label{homogen}\eqe

Stability analysis implies harmonic perturbations of the form $\exp(-i\omega t+ikx)$. In this case, Eqs. (\ref{cont}) and (\ref{Euler}) are reduced to
 \begin{eqnarray}
 -i\omega n_1+n_0\left(ikv_x+\frac{dv_z}{dz}\right)=0;\\
 \omega h_0v_x=kc^2p_1;\\
 -i\omega h_0v_z=-c^2\frac{dp_1}{dz}-gh_1.\label{vz}
 \end{eqnarray}
Eliminating velocities, one gets
 \eqb
\frac{d^2p_1}{dz^2}+\frac g{c^2}\frac{dh_1}{dz}-k^2p_1+\frac{\omega^2h_0}{c^2n_0}n_1=0.
 \eqe
One can express $p_1$ via $n_1$ making use of Eq.  (\ref{enthalpy}) and the thermodynamic identity $d(h/n)=dp/n$; this yields
 \eqb
p_1=(\Gamma-1)\left(\frac{h_0}{n_0}-mc^2\right)n_1.
 \label{p1}\eqe
Now one can write the closed equation for $n_1$:
 \eqb
\frac{\Gamma p_0}{n_0}\frac{d^2n_1}{dz^2}+
g\left(m+\frac{\Gamma^2}{\Gamma-1}\frac{p_0}{c^2n_0}\right)\frac{dn_1}{dz}
+\left(\frac{\omega^2}{c^2}\frac{h_0}{n_0}-k^2\frac{\Gamma p_0}{n_0}\right)n_1=0.
 \eqe
Note that we implicitly used the condition (\ref{homogen}) that the system is nearly homogeneous.

Looking for the solution in the form $n_1\propto\exp{\kappa z}$, one finds
\eqb
\kappa=-\frac g{2c^2}\left(\frac{\Gamma}{\Gamma-1}+\frac{mn_0c^2}{\Gamma p_0}\right)
\pm\sqrt{\frac{g^2}{4c^4}\left(\frac{\Gamma}{\Gamma-1}+\frac{mn_0c^2}{\Gamma p_0}\right)^2+k^2-\frac{h_0}{\Gamma p_0}\frac{\omega^2}{c^2}}.
\label{kappa}\eqe
The condition (\ref{homogen}) suggests that for $k\Delta\gtrsim 1$, one can neglect the terms with $g$ as compared with those with $k$. Qualitative considerations, which will be justified by the exact solution, show that the growth rate is determined by the free-fall time-scale, $\omega\sim\sqrt{kg}\ll k$, therefore one can also neglect the term with $\omega$. Then Eq. (\ref{kappa}) is reduced simply to $\kappa=\pm k$.

Taking into account that the perturbation should decay far enough from the slab, one can write
\begin{eqnarray}
p_1=C_1e^{kz};&\qquad z<0;\label{bot}\\
p_1=C_2e^{kz}+C_3e^{-kz};&\qquad 0<z<\Delta;\label{med}\\
p_1=C_4e^{-kz}&\qquad z>\Delta;\label{up}
\end{eqnarray}
where the constants $C$ could be found from the condition of continuity at the boundaries of the slab.

Let us consider the boundary $z=0$.  Defining the plasma shift, $\xi$, as $\partial\xi/\partial t=v_z$; i.e.
 \eqb
 \xi=i\frac{v_z}{\omega};
 \label{xi}\eqe
one can write the continuity of the flow and of the pressure as
 \begin{eqnarray}
 v_z(0)=v'_z(0);\label{cont_flow}\\
 p_0(\xi)+p_1(0)=p'_0(\xi)+p'_1(0);\label{cont_pressure}
 \end{eqnarray}
where the primed quantities are referred to the hot slab and the unprimed ones to the highly magnetized medium. For the wavelength $k\Delta\gtrsim 1$,  Eq. (\ref{vz}) is reduced, with account of the condition (\ref{homogen}), to
 \eqb
v_z=-\frac{1}{\omega h_0}\frac{dp_1}{dz}.
 \label{vz1}\eqe
Making use of Eq. (\ref{bot}) for $p_1$ and Eq. (\ref{med}) for $p'_1$, one writes Eq. (\ref{cont_flow}) as
 \eqb
\frac{1}{h_0}C_1+\frac{1}{h'_0}(C_2-C_3)=0.
 \eqe
The pressure continuity condition (\ref{cont_pressure}) is reduced to an equation for $C_1$, $C_2$ and $C_3$ if one takes into account that according to Eqs. (\ref{equil}), (\ref{xi}) and (\ref{vz1})
 \eqb
p_0(\xi)=p_0(0)+\frac{dp_0}{dz}\xi=p_0(0)-\frac{g}{c^2\omega^2}\frac{dp_1}{dz}.
 \eqe
Then one gets
 \eqb
\left[1-\frac{kg}{\omega^2}\left(1+\frac{h'_0}{h_0}\right)\right]C_1-C_2-C_3=0.
 \eqe

Applying the same considerations to the boundary $z=\Delta$,
one finds two more equations
 \begin{eqnarray}
 \frac 1{h'_0}\left(C_2e^{k\Delta}-C_3e^{-k\Delta}\right)+\frac 1{h_0}C_4=0;\\
 C_2e^{k\Delta}+C_3e^{-k\Delta}-C_4\left(1-\frac{g(h'_0-h_0)k}{h_0\omega^2}\right)=0.
 \end{eqnarray}
Now we have four linear homogeneous equations for the coefficients $C$. The solvability condition for this set of equations provides the dispersion equation. After simple algebra, one gets
 \eqb
\omega^4=k^2g^2\frac{1-\exp(-2k\Delta)}{\left(\frac{h_0+h'_0}{h_0-h'_0}\right)^2-\exp(-2k\Delta)}.
 \label{omega}\eqe
One sees that this equation always possesses an unstable root, ${\rm Im}\,\omega<0$. One can check that if the gravity force is directed downwards, $g>0$, the bottom boundary of the slab is unstable whereas $g<0$ implies instability of the upper boundary.

In the strongly magnetized medium, $h_0=2p_0$. In the relativistically hot slab, one can write $h'_0=4p'_0=4p_0$. Then one gets
\eqb
\frac{h_0+h'_0}{h_0-h'_0}=3.
\label{h}\eqe
In the small wavelength limit, $k\Delta\gg 1$, the growth rate could be found from equations (\ref{omega}) and (\ref{h}) as
\eqb
\eta\equiv -{\rm Im}\,\omega=\sqrt{\frac{kg}3},
\eqe
which coincides with the growth rate for the instability of the boundary between the two half-spaces, one filled with the plasma and another with the magnetic field. In the opposite limit, $k\Delta\ll 1$, one finds
 \eqb
\eta=\left(\frac g2\right)^{1/2}\Delta^{1/4}k^{3/4}.
 \eqe
In this regime, the plasma slab is bent as a whole.

\section{Interplay between magnetic dissipation and acceleration}

Now let as consider a highly relativistic outflow in which the magnetic field forms stripes of the opposite magnetic polarity containing cold plasma separated by currents sheets containing hot plasma. Let us denote the width of the stripe $l$.
In pulsar winds, $l$ is of the order of the light cylinder radius; specifically in the equatorial plane, $l=\pi r_L$. In accreting systems, $l$ could be as small as the gravitational radius, $l\sim r_g$.

When the flow accelerates, the effective gravity in the proper frame is
 \eqb
 g'=c^2\frac{d\gamma}{dr}.
 \label{g}\eqe
In this section, all quantities measured in the proper plasma frame are marked by primes.
As it was demonstrated in the previous section, the hot plasma drips out of the sheet in this case so that the slab eventually tears at some points (see Fig. 1) facilitating the magnetic reconnection. The most dangerous are perturbations with the wavelength comparable with the thickness of the slab.
Small wavelength perturbations just produce ripples on the slab surface whereas long wavelength perturbations result only in bending of the slab. Therefore the characteristic reconnection time is determined by the instability growth rate for the wavelength of the order of the sheet width,
 \eqb
\tau'\sim\sqrt{\Delta'/g'}.
 \eqe

As the field annihilates, the current sheet expands consuming the energy and the plasma from the magnetized domain. The rate of expansion could be written as
 \eqb
\frac{d\Delta'}{dt'}=\zeta\frac{\Delta'}{\tau'}=\zeta \sqrt{g'\Delta'},
 \label{reconn_rate}\eqe
where $\zeta$ is a numerical factor. Three-dimensional numerical simulations of the current sheet supported against the gravity force are necessary in order to confirm this scaling and to provide an estimate for $\zeta$.

Substituting equation ({\ref{g}}) into equation (\ref{reconn_rate}) and transforming to the lab frame, $r=ct'\gamma$, $\Delta=\Delta'/\gamma$, one can write the rate of the sheet expansion as
 \eqb
\frac{d\gamma\Delta}{dr}=\zeta\sqrt{\gamma\Delta\frac{d\gamma}{dr}}.
 \label{expansion_rate}\eqe
An important point is that dissipation of the Poynting flux forces the flow to accelerate  (Lyubarsky \& Kirk 2001; Drenkhahn 2002; Drenkhahn \& Spruit 2002; Kirk \& Skj{\ae}raasen 2003) therefore the process of dissipation due to the Kruskal-Schwarzschild instability is self-sustaining: the acceleration facilitates the dissipation whereas the dissipation leads to the acceleration. At the initial stage, Poynting dominated flows are accelerated even without dissipation therefore the process could be initialized naturally.
Below we present the self-consistent description of the acceleration/dissipation process making use of the short-wavelength approximation, $l\ll r$, developed by Lyubarsky \& Kirk (2001).

We consider the radial flow with a purely toroidal magnetic field. This approximation is valid far enough from the source where
one can neglect the poloidal field in the proper frame of the flow,
 \eqb
 \gamma\gg B_{\phi}/B_p=R/r_L.
  \label{toroidal_cond}\eqe
Here $R$ is the cylindrical radius of the flow.

The flow is steady on the average therefore
the continuity and the energy conservation equations could be written as
 \begin{eqnarray}
r^2 \langle nv\rangle=\it const;\\
 r^2\left\langle h'\gamma^2v+\frac{B^2}{4\pi}v\right\rangle=\it const;
 \end{eqnarray}
where angle brackets denote averaging over the striped structure. Adopting for simplicity the rectangular stripe shape such that within the current sheet, the magnetic field is zero and the plasma is relativistically hot with the pressure balancing the magnetic pressure of the cold, magnetized domain, $h'=4p=B'^2/2\pi$, one writes the conservation equations as
 \begin{eqnarray}
 vr^2[n_c\Delta+(l-\Delta)n_h]=\it const;\\
 v\gamma^2r^2\left[\frac{B'^2}{2\pi}\Delta+(l-\Delta)\left(mn'_cc^2+\frac{B'^2}{4\pi}\right)\right]
 =\it const.
 \end{eqnarray}
Here $n_c$ and $n_h$ are the plasma density in the cold magnetic domain and in the hot current sheet, respectively. The constants are determined by the parameters at the inlet of the flow, $r=r_0$, which should be chosen far enough from the central source so that the condition (\ref{toroidal_cond}) is already fulfilled.

In the limit $\Delta\ll l$, $\gamma\gg 1$, these equations may be combined into the asymptotic relation (equation (29) in Lyubarsky \& Kirk 2001; note the change in notations: Lyubarsky \& Kirk used dimensionless $\Delta$, which is equal to our $\Delta/l$)
 \eqb
\sigma_0\left[\left(2\frac{n_h}{n_c}-3\right)\frac{\Delta}l+\frac 1{2\gamma_0}-\frac 1{2\gamma^2}\right]=\frac{\gamma}{\gamma_0}-1;
 \label{asymp}\eqe
where $\gamma_0$ is the Lorentz factor at the inlet of the flow,
 \eqb
\sigma_0=\left(\frac{B^2}{4\pi mn_c\gamma c^2}\right)_{r=r_0}
 \eqe
the initial magnetization parameter.
This equation should be complemented by the entropy equation and by the prescription for the reconnection rate.  In our case, the last is given  by equation (\ref{expansion_rate}). The entropy equation implies $n_h=3n_c$ at large distances (Lyubarsky \& Kirk 2001).

When $\Delta/l\gg 1/\gamma_0^2$, equation (\ref{asymp}) is reduced to
 \eqb
3\sigma_0\frac{\Delta}l= \frac{\gamma}{\gamma_0}.
 \label{Delta}\eqe
Substituting this relation into equation (\ref{expansion_rate}), one gets the equation for $\gamma$
 \eqb
\gamma^2\frac{d\gamma}{dr}=\frac{3\zeta^2\sigma_0\gamma_0}{4l},
 \eqe
which is immediately solved to yield
 \eqb
\gamma=\left(\frac{9\zeta^2\gamma_{\rm max}r}{4l}\right)^{1/3};
 \eqe
where $\gamma_{\rm max}=\sigma_0\gamma_0$ is the Lorentz factor achieved if the Poynting flux is completely transformed into the plasma kinetic energy. Combining this relation with
equation (\ref{Delta}) one sees that the Poynting flux is dissipated completely, $\Delta\sim l$, at the scale
 \eqb
r_{\rm diss}=12\left(\frac{\gamma_{\rm max}}{\zeta}\right)^2l.
 \label{diss_radius}\eqe
At this scale, $\gamma$ approaches $\gamma_{\rm max}$.

\section{Discussion}

In this Letter, we proposed a novel mechanism for the energy release in Poynting dominated outflows. The mechanism operates if a significant fraction of the Poynting flux is transferred by alternating fields. Such a structure inevitably arises in the wind from obliquely rotating pulsars and magnetars. In accreting sources, one has to assume that the magnetic polarity changes sign near the black holes.
Earlier only anomalous resistivity mechanisms for the field dissipation were considered; they are  efficient only in the current starvation regime when the thickness of the current sheet is comparable with the Larmor radius of the particles (Coroniti 1990; Lyubarsky \& Kirk 2001; Kirk \& Skj{\ae}raasen 2003; Zenitani \& Hoshino 2007). We found a universal mechanism for dissipation of alternating fields, which works even if the conditions of ideal MHD are fulfilled and anomalous resistivity does not appear. Thus we justified the assumption of Spruit et al. (2001) that in Poynting dominated outflows, the annihilation rate of alternating fields comprises a good fraction of the speed of light.

The mechanism is based on the MHD instability of the current sheet in the accelerating flow. Under the influence of the inertia force, the plasma drips out of the current sheet allowing the oppositely directed fields come together. The ensuing field annihilation leads to the flow acceleration so that the process is self-sustaining. Note that this instability is easily suppressed by magnetic shear therefore it is very important that in the far zone of the flow, the field becomes predominantly toroidal, i.e. practically shearless.

We have shown that the complete dissipation occurs at the scale (\ref{diss_radius}). In pulsar winds, $\gamma_{\rm max}\sim 10^4-10^6$ and $l\sim r_L$ so that with this mechanism, the alternating fields are dissipated at the distance  $\sim 10^9-10^{13}r_L$, which is comparable or even larger than the dissipation scale due to the current starvation mechanisms (Lyubarsky \& Kirk 2001; Kirk \& Skj{\ae}raasen 2003; Zenitani \& Hoshino 2007).
However, this mechanism could efficiently work in GRBs and AGNs because $\gamma_{\rm max}$ in these systems is not very large whereas the current starvation distance is huge.

In AGNs, $\gamma_{\rm max}\sim 10$ and $l\sim r_g$ therefore this mechanism provides the energy release at the scale of $\sim 1000r_g$. This is compatible with observations because fitting of the blazar spectra implies that jets are already matter dominated at $\sim 1000 r_g$  (Ghisellini et al. 2010). The proposed mechanism provides also a natural basis for the "jets-in-a-jet" model (Giannios et al. 2009, 2010) that accounts for the observed ultra-fast TeV variability of blazars.

In GRBs, $\gamma_{\rm max}\sim 1000$ so that the energy release occurs at the distance of the order of $10^{13}$ cm. The magnetic dissipation at this scale could reproduce the observed properties of the prompt GRB emission (Giannios \& Spruit 2007; Giannios 2008). High efficiency of the conversion of the magnetic energy into the energy of radiating particles makes this model a viable alternative to the internal shock model.

\begin{acknowledgments}
This work was supported by the US-Israeli Binational Science Foundation under grant number 2006170 and by the Israeli Science Foundation under grant number 737/07.
\end{acknowledgments}

\hyphenation{Post-Script Sprin-ger}

 \end{document}